\documentclass[runningheads]{llncs}
\usepackage{graphicx}
\usepackage{pgfplots}
\usepackage{hyperref}
\usepackage{textcomp}
\usepackage{listings}
\usepackage{amsmath}
\usepackage{amssymb}
\usepackage{color}
\usepackage{forest}
\usepackage{dsfont}
\usepackage{import}
\usepackage[pdf]{pstricks}
\usepackage[
	n,
	operators,
	advantage,
	sets,
	adversary,
	landau,
	probability,
	notions,
	logic,
	ff,
	mm,
	primitives,
	events,
	complexity,
	asymptotics,
	keys]{cryptocode}
\hypersetup{linkcolor=black, pdfborder={0 0 0}}

\definecolor{mygreen}{rgb}{0,0.6,0}
\definecolor{mygray}{rgb}{0.5,0.5,0.5}
\definecolor{mymauve}{rgb}{0.58,0,0.82}

\newif\iflongversion

\newcommand{\ie}{i.e., }%
\newcommand{\queries}{\mathcal{Q}}
\newcommand{\policies}{\Phi}

\newcounter{Att}
\newcommand{\att}[1]{%
\customlabel{#1}{A.\theAtt}%
(A.\theAtt)%
\stepcounter{Att}%
}
\newcounter{Req}
\newcommand{\req}[1]{%
\customlabel{#1}{R.\theReq}%
(R.\theReq)%
\stepcounter{Req}%
}
\makeatother
\newcommand{\inlineEnum}[1]{%
  \ifcsname c@#1\endcsname%
  \addtocounter{#1}{1}%
\textrm{\arabic{#1})}~%
\else%
\newcounter{#1}%
\setcounter{#1}{1}%
    \textrm{\arabic{#1})}~%
    \fi%
}
\makeatletter
\newcommand{\customlabel}[2]{%
  \protected@write\@auxout{}{\string\newlabel{#1}{{#2}{\thepage}{#1}{page.\thepage}{}}}%
}

\begin{document}
\title{Data Querying and Access Control for Secure Multiparty Computation\thanks{This work has been supported by the German Federal Ministry of Education and  Research,  project  DecADe,  grant  16KIS0538  and  the  German-French Academy for the Industry of the Future.}}

\author{Marcel von Maltitz  \and
Dominik Bitzer \and
Georg Carle}
\authorrunning{Marcel von Maltitz et al.}
\institute{Chair for Network Architectures and Services\\ Department of Informatics, Technical University of Munich \\
  \email{\{lastname\}@net.in.tum.de}\\
}
\maketitle              
\begin{abstract}
In the Internet of Things and smart environments data, collected from distributed sensors, is typically stored and processed by a central middleware.
This allows applications to query the data they need for providing further services.
However, centralization of data causes several privacy threats:
The middleware becomes a third party which has to be trusted,
linkage and correlation of data from different context becomes possible and
data subject lose control over their data.

Hence, other approaches than centralized processing should be considered.
Here, Secure Multiparty Computation is a promising candidate for secure and privacy-preserving computation
happening close to the sources of the data.

In order to make SMC fit for application in these contexts, we extend SMC to act as a service:
We provide elements which allow third parties to query computed data from a group of peers performing SMC.
Furthermore, we establish fine-granular access control on the level of individual data queries, yielding data protection of the computed results.
By adding measures to inform data sources about requests and the usage of their data, we show how a fully privacy-preserving service can be built on the foundation of SMC.

%

\keywords{Privacy \and
Transparency \and
Intervenability \and
Access control \and
Internet of Things \and
Secure Multiparty Computation.}
\end{abstract}
\longversiontrue
\longversionfalse

\section{Introduction}
In the last years, the Internet of Things (IoT) has become an emerging trend,
including the  rise of smart environments.
\iflongversion
These contexts are equipped with a multitude of sensors to make deployed IT systems aware of the physical state of the environment.
Based on this data, services are provided which interact with users or influence the environment through actuators.
\cite{RAOE}

Here, a certain structure of applications is prevalent:
\else
In these contexts, a certain structure of applications is prevalent:
\fi
Data points are collected at different (spatial or logical) locations;
for their utilization, they are horizontally aggregated over a set of collectors; the outcomes are statistics or higher-level information which are then used by other services.
The corresponding architecture normally features sensors for data collection, a middleware for data storage and processing and services which receive the processed data for further usage.
\iflongversion
Examples for services are public displays that provide aggregated data about the environment's state to the users, controllers which are triggered by data change events or even machine learning systems, where the single data points are incorporated into a collective model.
\else
Examples for services are public displays providing statistics and controllers or actuators triggered by data change events.
\fi

However, the data which is initially collected by sensors is privacy-sensitive since it typically reflects user presence and interaction:
Presence influences the temperature in single rooms, the $\mathrm{CO}_2$ concentration and device utilization (e.g. lighting).
Certain sensors, e.g. for power consumption tracking provide even more specific insights into user behavior (Intrusive Load Monitoring, cf. \cite{ILMSurvey}).
This raises a huge need for information security and privacy, which is not well addressed by centralized storage and processing solutions.

Secure Multiparty Computation is an emerging and highly promising approach for providing privacy-preserving systems.
It encompasses a class of protocols which allows a group of parties to compute arbitrary functions based on confidential inputs without sharing these values with any other party.
With the advent of edge computing \cite{EdgeComputingVision}\!\cite{EmergenceEdge} it also becomes a promising alternative in the context of smart environments:
Data can remain on local sensor platforms in the proximity of the sensors where it was collected.
Execution of SMC protocols between these nodes then enables secure processing without sharing or centralizing raw sensor data.
\iflongversion
This solves the conflict between the privacy-criticality of the individual data points versus the utility of results derived from them by further processing.
\fi

\iflongversion
However, SMC is not directly applicable in the depicted context:
Typically, SMC is carried out in peer-to-peer like infrastructures (cf.~\cite{BGW88,BristolMPC}): By default the nodes provide the input, perform the computations and receive the results themselves.
Alternatively, the set of computing parties is reduced to a small number of dedicated hosts, preserving some kind of centrality.
The data of smart environments, however, is initially distributed.
Hence, it  is beneficial to exploit this distribution by performing SMC on nodes in the proximity of the respective sensors.
Furthermore, these contexts demand service-oriented architectures:
\else
However, SMC is not directly applicable in the depicted context due to an architectural mismatch: SMC uses a peer-to-peer like communication (cf.~\cite{BGW88,BristolMPC}) while smart environments are service-oriented:
\fi
Nodes provide services for other clients; in particular, clients should be able to request specific data processing and receive the result afterwards.
Our goal in this paper is to extend SMC to be applicable in the architecture of smart environments. Using a state of the art notion of privacy, we yield a fully privacy-preserving service for the processing  of sensor data.

The remainder of this paper is structured as follows:
In Section~\ref{sec:background} we provide our notion of privacy and a background on SMC.
In Section~\ref{sec:related_work}, we elaborate the related work on practical application of SMC.
Section~\ref{sec:goal_statement} provides a goal statement of our work.
We present our approach in Section~\ref{sec:approach}
while Section~\ref{sec:discussion} discusses the achievements afterwards.
A performance evaluation of our approach is carried out in Section~\ref{sec:evaluation}.
Section~\ref{sec:summary} concludes the paper.
\section{Background}
\label{sec:background}
In this section, we provide the background to our notion of privacy, SMC in general and the Security and Privacy Model of SMC applied to our domain of smart environments.

\subsection{Privacy}
\label{sec:background_privacy}
Initially, privacy and data protection are high-level concepts\iflongversion{} that possess a broad horizon of meaning and interpretation\fi{}.
In order to realize these properties in an information technological system, it is necessary to further refine them and break them down into more specific parts.

\iflongversion
The method of defining protection goals has already proven beneficial for the concept of security:
Its dissection into \emph{confidentiality}, \emph{integrity} and \emph{availability} gave protection goals at hand which can be
achieved and verified in information systems.
Today, they constitute a common and fundamental understanding of information security.

In the last years, privacy and data protection underwent a similar process:
More specific protection goals were formulated that should make the concepts more graspable.
\else
Like security years before, privacy and data protection recently underwent the process of refinement by defining protection goals, which make the concepts more graspable.
\fi
Following \cite{Datenschutzschutzziele,Pfitzmann2010,Hansen2015,OrdnungPrivacySchutzziele}, these protection goals mainly are
\emph{data minimization},
\emph{unlinkability},
\emph{transparency} and
\emph{intervenability}.
\iflongversion
Here, data minimization declares that no data should be created or collected if it is not specifically needed or the necessity is not clear beforehand.
Unlinkability means that it should not be possible to combine available data in a manner to generate more insights than initially intended.
This can either be different data of a single entity or data of several entities, that is used in order to derive information about interactions or correlations.
Transparency is required with respect to the information processing systems.
Data subjects, the entities about whom the data is collected, should be able to understand how and why their data is processed by whom.
This is the foundation for intervenability, requiring that data subjects remain in control of their data although it is processed by a third party.
In general, this includes revocation of access, correction and deletion of data.
\else
For the standard definition we refer to the mentioned literature, especially \cite{Datenschutzschutzziele}.
An interpretation in our specific context is provided in Section~\ref{sec:sec_model}.
\fi

We utilize this notion of privacy and data protection, since it is state of the art and has found widespread adaptation, be it on the level of individual states~\cite{SDM,WhiteHouseReport} or the European Union~\cite{ENISA2014}.

\subsection{Secure Multiparty Computation (SMC)}
SMC formalizes a problem of controlled leakage.
Assume multiple cooperating parties, each holding a confidential value.
They agree on some function which takes these values as inputs.
Using SMC, the function is correctly evaluated while its result is the only new information released.
The input of each peer is not shared with any third party including the other cooperating peers.  \cite{SMCBook}

The seminal work of Yao \cite{Yao1982,Yao1986} laid the foundation of SMC; from there, several different methods for realization emerged
(e.g., \cite{BGW88,Chaum1988,Goldreich1987}).
The most promising foundations for SMC currently are garbled circuits, homomorphic encryption and secret sharing schemes.
These approaches enable different {usage models} \cite{Archer2016}, i.e.,
outsourced processing, outsourced services and joint processing.
For a comprehensive overview see \cite{Archer2016}.
Today, research mainly focuses on the performance of general purpose computation suites, strengthening their security, identifying new fields of application and designing efficient specific purpose protocols.
\cite{BristolMPC,BreakingSPDZ,MASCOT,Keller2018}

With regard to privacy protection, SMC naturally fulfills data minimization and unlinkability.
Data minimization is provided since raw data can remain where it is created.
Desired results can be computed by SMC directly without creating privacy-critical intermediary data.
Two types of unlinkability are given:
By aggregation\iflongversion\footnote{Reducing the information density by aggregation is an established privacy design pattern \cite{Hoepman2014}.}\fi{} of multiple parties' input data, linkability between the individual input data and the result is prevented.
Tracing back unique parties' inputs from a result is impossible in the general case.
Furthermore, linkability among different parties, i.e., correlation of their data, is prevented, since this data is never available at the same logical location.

Currently, a small amount of actively developed SMC frameworks \cite{frescoURL,BristolURL,mambaURL} exists.
These provide an implementation of the basic operations and enable creation of arbitrary composed algorithms.
Their application is initially restricted to the distributed execution of the created protocols, which does not encompass management and orchestration of peers, coordination of computations
and infrastructural requirements enabling application of SMC in data processing scenarios.

\subsection{Security and Privacy Model}
\label{sec:sec_model}
\paragraph{Assets}
The main asset to be protected is the individual raw data of sensor platforms.
We assume them to be owned each by the respective data subject, i.e., the person(s) about
which the platform gathers information.
This is given in use cases where a single smart building is inhabited by different parties, e.g., smart hotels, smart houses with individual rental apartments
and can also be given in smart office buildings, if employees have dedicated offices.
The necessity for data protection is based on the possibility that sensor data gives insights about the presence and behavior of individuals \cite{Roman2013}.
\paragraph{Protection Goals}
\label{sec:protection_goals}
With respect to the mentioned assets we understand security to be confidentiality of this very raw data.
However, confidentiality may not hinder all processing of the data.
Instead, a privacy-preserving access must be designed, meaning data access which is controllable by and accountable for the data owners.
Following our privacy background in Section~\ref{sec:background_privacy},  privacy-preservation encompasses the protection goals of \inlineEnum{pp} data minimization, \inlineEnum{pp} unlinkability, \inlineEnum{pp} transparency, and \inlineEnum{pp} intervenability.
They have the following meaning in our context:
\inlineEnum{ppa} Information is only derived from raw data if it is actually needed by any client service.
The purpose is known before information is created.
\inlineEnum{ppa} Information made available to clients does not allow restoring contributions of individual single peers.
Correlations between individual peers should not be possible by client accessible data.
\inlineEnum{ppa} Peers should know, what information is derived from their data and for which purpose this information is used.
\inlineEnum{ppa} Based on this preliminary knowledge, they should remain in control of their data by deciding which computations may be carried out.
\paragraph{Attacks}
\iflongversion
In classical architectures, raw sensor data is pushed from sensor platforms to a centralized middleware.
There, data analysis and processing is performed.
The results are then made available to clients.

In this architecture,
\else
In classical architectures
\fi
previously mentioned protection goals are not or only partially fulfilled.
\iflongversion
Confidentiality is not given, since raw data is forwarded to and stored on a third party middleware.
The middleware becomes a single point of attack and a high value target.
Data minimization is not enforced since arbitrary derivations of information can be performed after its collection.
Concomitantly, the purpose of data is not controllable after collection, i.e., using the same data for other purposes is possible at any time in the future.
Especially, new information can be generated by correlating raw data of different sources.
All this is possible without the individuals knowing and enabling them to intervene in the processes.
\else
Central storage creates a high value target and a single point of attack.
Furthermore, it allows arbitrary processing, using data for other purposes and correlation of available data; all this being completely intransparent for individuals and without any ability to intervene.
\fi

In our architecture, these attacks are mitigated, since raw data stays on the peers,
clients only obtain the post-processed information they have been granted access beforehand.
That information is generated by SMC in a privacy-preserving manner and the gateway only orchestrates data processing while not having access to
raw data\iflongversion{} as a classical middleware would\fi{}.

Introducing these three roles, further attacks have to be considered:
Malicious peers can try to obtain information from other peers \att{att:peercurious} or provide wrong results \att{att:peercorrectness}.
A malicious gateway can try to obtain information on their own behalf \att{att:gatewayrequest}, collect \att{att:gatewaycurious} or tamper with results \att{att:gatewaytamper} obtained from peers before forwarding.
Malicious clients can try to obtain information without having the appropriate access right \att{att:clientaccess} or correlate \att{att:clientcorrelate} information they were able to obtain.

In our considerations, we exclude \ref{att:peercurious} since it has to be addressed on the level of the SMC protocols and \ref{att:peercorrectness} since the correctness of input values provided
by peers is out of scope of realizing secure computation (cf. \cite[p.~11]{SMCBook}).
Furthermore, \ref{att:clientcorrelate} is excluded since it depends on the exact choice of available computation queries.
While designing our approach to enable SMC usage, we consider the remaining abovementioned attacks \ref{att:gatewayrequest}--\ref{att:clientaccess}.

\paragraph{Trust}
An ultimate design goal is to reduce the amount of components which have to be trusted to handle private raw data faithfully.
Our architecture has been designed to avoid single points of attack and high value targets holding private raw data from several parties.
The remaining trust is diversified:
We associate each sensor platform with individual users.
These users trust their respective platform to faithfully collect and store their data.
This assumption does not strongly differ from assumptions in classical architectures:
In any case, by generating it, sensors have access to privacy-critical data.
Furthermore, all trust requirements for the used SMC protocol realization apply.

\section{Related Work}
\label{sec:related_work}
Several results were achieved in the last years.
However, they show mere feasibility without aiming for an automated system providing SMC as a service.
In \cite{SugarBeetSMC} SMC was used to perform an auction between buyers and sellers of a specific product.
Providing data and receiving results was executed manually.
Another auction was performed in \cite{SMCAviation} among different airlines
for implementing the EU Emission Trading Scheme.
Data input and output are performed using CSV files.
In \cite{SMCFinancial} a comparison of key performance indicators among a group of competitive companies was performed.
They provided a Javascript library enabling data collection via the browser; computation results were made available via a spreadsheet.
Burkhart et at. \cite{SEPIANetwork} applied SMC to generate network traffic statistics and anomaly detection. Similarly, \cite{SMCOutage} reused the same framework to perform collaborative outage detection. Both do not address deployment and data access challenges.
Recently, Bonawitz et al. \cite{Bonawitz2017} used SMC to collect private user data from smartphones in order to train a central machine learning model.
For this specific use case, they provide a solution which is intended to serve as an automated service collecting the desired data.

All of these solutions fall short for the application in the IoT.
In most cases, data is provided manually by user interaction.
Similarly, the SMC setup is created ad-hoc for single computations\iflongversion, instead of a continuous deployment for automated requests\fi.
Correspondingly, computations are invoked by manual intervention.
Also, the architecture does not match:
Data providers and result consumers are the same entities.
They cooperate in a peer-to-peer fashion processing data for themselves instead of providing a service for third parties.

A notable difference is the last mentioned work.
They actually provide an automated SMC service.
However, while optimizing for a certain use case,
they sacrifice the ability to compute arbitrary functions and specialize on
secure aggregation.
Following \cite{vonMaltitz2018b}, our solution is agnostic regarding the specific SMC implementation as long as it supports an arbitrary number of computing parties.

\subsection{Previous work for SMC in the Internet of Things}
\label{sec:previous_smc_iot}
In \cite{vonMaltitz2018c} von Maltitz et al. provide a vision how SMC can be applied in smart environments:
The starting point are distributed sensor platforms, understood as edge devices.
They represent an intermediary to low-end sensors; they collect the data created by the connected sensors, store it locally and have sufficient resources to perform local, small-sized data processing.
SMC computations among these devices allow to derive processed and aggregated information from this local data, which is then made available so that services (public displays, actuators, etc.) can act on them.

As middleware, an \emph{SMC gateway} is deployed.
It realizes the link between the sensor platforms and the data consuming services but without having access to or storing the sensor data.
Regarding the sensor platforms -- called \emph{peers} --,  it coordinates SMC sessions for data processing.
Towards the services -- called \emph{clients} --, the SMC gateway acts as a middleware which allows querying of data in the standard client server paradigm while abstracting from application of SMC.

In \cite{vonMaltitz2018b} von Maltitz et al. focused on the interaction between the peers and the SMC gateway of the aforementioned vision.
This work addressed natural technical mismatches between the premises of secret sharing based SMC~\cite{BGW88,Shamir1979} and the characteristics of dynamic environments.
The result is a management and orchestration framework for SMC which
\iflongversion
specifically provides
\inlineEnum{challenges}auto-discovery of potential peers,
\inlineEnum{challenges}adaptivity regarding the data made available by new peers,
\inlineEnum{challenges}orchestration of peers during SMC sessions and
\inlineEnum{challenges}error recovery caused by vanishing peers or interrupted connections.
This work
\fi
enables stable and automated execution of SMC sessions in dynamic environments.

However, since it only provides the first purpose of the SMC gateway, the remaining challenges for extending SMC, \ie enabling data querying and access control on data processed by SMC,  are addressed in this work.

\section{Open Problems and Goal Statement}

\label{sec:goal_statement}
The first goal of this work is to allow clients to query data from a coordinated SMC group.
While doing so, the fact that SMC is used for data processing should be abstracted away for the requesting clients.
For this, we need queries which describe the results to be obtained in a non-SMC specific manner \req{req:queries}.  
Protection of original sensor data is achieved by SMC.
They never leave the sensor platforms and are not shared with any third party.
However, access control has to be carried out on the derived results, ensuring that no client obtains more or other data than intended \req{req:accesscontrol}.

The second goal is a provide a fully privacy-preserving service based on SMC.
This means in addition to fulfilling unlinkability using SMC, transparency \req{req:transparency}, including accountability \req{req:accountability},  and intervenability \req{req:intervenability} should be achieved according to their definitions in Section~\ref{sec:sec_model}.

\section{Approach}
\label{sec:approach}
We extend the gateway solution of \cite{vonMaltitz2018b}.
First, we define a format for \emph{queries}.
A query is a declarative data structure which stands for a computation the gateway offers and clients can ask for.
Corresponding \emph{authorization grants} are specified that state the permission which queries a client is allowed to post.
Lastly, three types of requests are defined:
\emph{Metadata requests} request the set of available queries from the gateway.
\emph{Grant requests} are sent from the client to the gateway in order to obtain an authorization grant, permitting a certain query.
\emph{Computation requests} are actual requests for data, specifying the data to be obtained using a query and providing a corresponding grant authorizing them to obtain this very data.

\begin{figure}
  \includegraphics[width=\columnwidth]{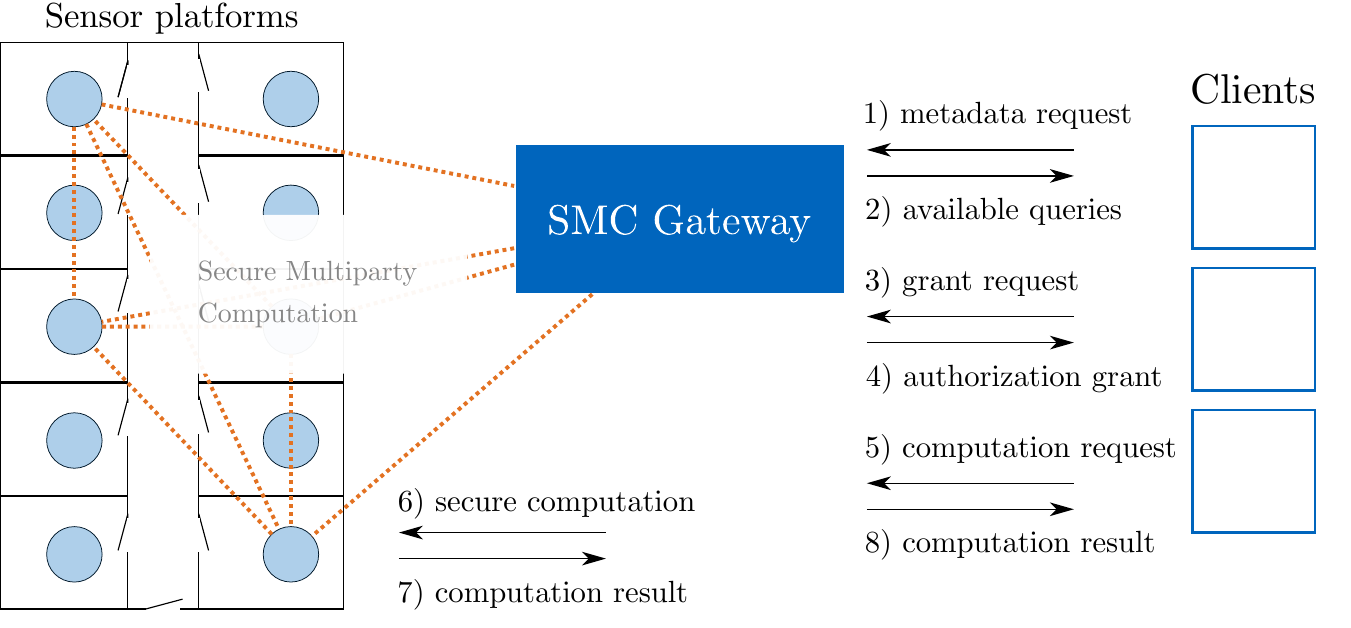}
  \caption{Interactions between the clients, the SMC gateway and the SMC peers. }
  \label{fig:arch}
\end{figure}

The subsequent protocols enable the following interaction (cf. Figure~\ref{fig:arch}):
Clients perform a metadata request to the gateway (1),  obtaining meta information about all available queries via this gateway (2).
From this set clients select desirable queries and ask for permission to issue them by performing a corresponding grant request (3).
On success, they obtain the respective authorization grant (4).
Data is demanded by a computation request (5).
When the gateway obtains this type of request, permission is checked using the authorization grant which is sent along with the request.
On success, the affected peers are informed, enabling them to also check and verify the request using the provided authorization grant.
Additionally, they can perform arbitrary further privacy checks.
If all peers agree, the SMC session is setup and carried out (6).
When the gateway receives the result (7), it is forwarded to the client (8).

\subsection{Requests and Authorization}
\label{sec:requests}
\paragraph{Cryptographic identity for clients}
Given the previous work of \cite{vonMaltitz2018b} we can already assume the availability of a public key infrastructure.
Clients have to be equipped with a cryptographic identity, e.g. X.509 certificates, for providing a secure reference point
for communication and authorizations to be bound.
In order to support transparency later on, the identity is enhanced with metadata about the clients, especially a short description which
states the usage purpose of data obtained by this client.
\iflongversion
Similarly, if applicable in the context of use, clearance information per client can also be provided.
\fi

\paragraph{Query format}
\label{sec:query_format}
For performing a secure computation the following properties of the computation must be known:
\inlineEnum{smcprop}The group of peers to participate in the computation
\inlineEnum{smcprop}the input data to use and
\inlineEnum{smcprop}the protocol to execute.

Regarding \inlineEnum{smcpropDiscussed}we refrain from letting clients specifically select single peers to form a group.
In our use case, the knowledge which is of interest for clients is on a higher abstraction level like a department, a floor, a specific room type etc.
\iflongversion
Hence, they should not have the burden of selecting the matching peers individually.
Instead,
\else
Hence,
\fi
this abstraction is made on the side of the gateway:
Newly added peers provide metadata about themselves, including labels describing peers in a domain-specific way.
We hence define $labels(p)$ to be the set of key-value pairs $(k,v)$ of peer $p$.
As an example, a peer can have the label set \texttt{\{roomtype: kitchen, level:3, buildingpart:A,\ldots\}}.

When a set $P_g$ of multiple peers is connected to a gateway $g$, it can build the superset $ L_{P_g} \equiv \bigcup_{p \in P_g} labels(p)$
of all provided attributes.
This information can then be used to create predefined logical predicates forming groups, which can then be queried by clients.
\iflongversion
Automatically, simple predicates can be derived e.g. by enumerating available attribute combinations:
$$ \forall (k,v) \in L_{P_g} : \texttt{``k = v''} \equiv  \{ p \in P_g  | (k, v) \in labels(p)\}$$
\fi
A predicate \texttt{``\text{roomtype} = \text{kitchen}''} would hence select all peers which communicated this label upon pairing with the gateway.
\iflongversion
Alternatively, manual predicates can be created, specifically forming semantically sensible and useful groups.
\fi
Clients can then choose from the finite set of predicates for each of their requests.
In order to guarantees privacy towards the peers, clients should not be allowed to craft queries themselves instead of selecting from a predefined set:
Making more general queries than provided could enable them to gain information about peers which they are not allowed to obtain.
Making them more specific could allow derivation of input data of single or a small group of peers.

Similar to the labels, for \inlineEnum{smcpropDiscussed}each peer provides metadata about the inputs it can provide.
We denote this as $inputs(p)$ for peer $p$.
It corresponds with the sensors the peer has available.
Besides the type of input data, clients are given the ability to select not only the latest data point but also a list of the points of a given time window reaching into the past.
I.e. as $preselector$ the gateway allows to choose from window sizes like \textit{last value}, \textit{last hour}, \textit{last 6 hours}, \ldots

When selecting a window of values, they have to be merged into a single value as input for the SMC session.
For this, we define a $preprocessor$ function that can be selected by the client and request a corresponding aggregation of the values before performing the actual computation.
These encompass typical aggregation functions like $min, max, sum, average, \ldots$ but can also arbitrarily extended.
\iflongversion
This preprocessing does not need any special privacy-protection since computation happens on each peer individually only with own data.
\fi

Regarding \inlineEnum{smcpropDiscussed}we depend on the protocols being available on the peers.
These can be provided as labels and clients can again select from a finite set of options.

In summary, the three requirements for SMC sessions are transformed into five attributes a request has to provide.
An example is given in Listing~\ref{lst:client_request}.
Request translation then conceptually consists of two steps:
\inlineEnum{reqtrans}The selected group label is evaluated and the corresponding peers are chosen.
All further request attributes are evaluated at the peers, selecting the right (preprocessed) input and the protocol to execute.
\iflongversion
Based on the data provided by the peers during pairing with the gateway, the gateway also informs each peer about all other
participants (cf. \cite{vonMaltitz2018b}) in order to enable SMC connections among them.
\else
The session is then carried out according to  \cite{vonMaltitz2018b}.
\fi

\begin{lstlisting}[escapechar=\+,caption=Computation request query, label=lst:client_request]
{
 predicate: type = heater +$\wedge$+
     roomtype +$\in$+ [kitchen, meetingroom]
 preselector: last 6 hours
 preprocessor: avg
 protocol: sum
 input: power_consumption
}
\end{lstlisting}

\paragraph{Authorization grants}
\iflongversion
A straightforward approach of access control is that the gateway represents the policy decision and enforcement point.
Depending on an internal state holding the current access rules, the gateway allows or rejects an incoming query.
Alternatively, the decision point is externalized into an own component which is then contacted by the gateway upon request.

These approaches have several disadvantages:
We focus on dynamic environments, i.e. neither continuous connectivity should be assumed and nor that the gateway always stays the same entity.
In consequence, permissions should neither depend on continuous connectivity to third party entities nor on extensive internal state of the gateway.
Furthermore, following our requirement of transparency and intervenability, requests must also be verifiable on the peers.
This requires that authorization information must also be accessible for them.

\else
In a dynamic environment, it is imaginable that multiple gateways are deployed for different purposes.
These gateways should be able to verify access requests without time or communication overhead.
I.e. no contact to other entities should be necessary nor should extensive state on the gateway itself be necessary for verification.
\fi
Due to these reasons, we decide for a serialized representation of permission which is transferable and verifiable with a low amount of state information.
The purpose of such an \emph{authorization grant} is to state whether a given request is legitimate or not.
The client can obtain these grants from an access authority (e.g. the gateway or an external entity).
These grants encode the requests allowed for this client.
Upon request, the clients also send the grant along in order to prove legitimacy of the request.
Forwarding the grant to the peers enables individual revalidation by the data sources themselves.

\iflongversion
In our case, the access authority is the gateway.
However, it can be externalized to another entity.
This is especially beneficial to reduce the necessary amount of trust in the gateway:
While compromising the access authority would allow creation of grants which would be considered valid by the peers,
a malicious gateway can only forward arbitrary requests, where the peers could still correctly distinguish between authorized and unauthorized requests.
\fi

Technically, the grant mirrors the attributes of the request.
I.e., it also contains a predicate which can be matched against the request predicate.
Furthermore, preselector, preprocessor and the input type must be of permitted value.
Besides that, it is bound to the identity of the corresponding client by adding a cryptographic identifier bound to the client's certificate.
To avoid complications regarding revocation, we suggest a short lifetime of the grants and renewal on demand.

\begin{figure}[t]
\begin{center}
\begin{forest}
  [grant request[certificate[fpr\\\ldots, align=left]][queries[query[predicate\\ preselector\\ preprocessing, align=left][ protocol\\ input, align=left]][\ldots][query]][$\mathrm{sig}_{\mathrm{client}}$]]
\end{forest}
\end{center}
\caption{Structure of a grant request}
\label{fig:grant_query}
\end{figure}
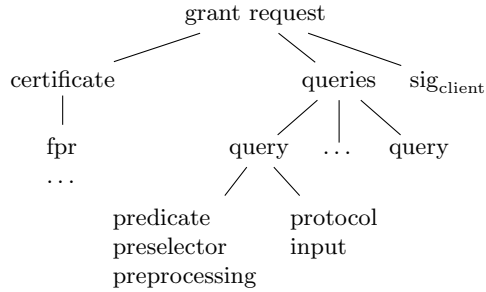

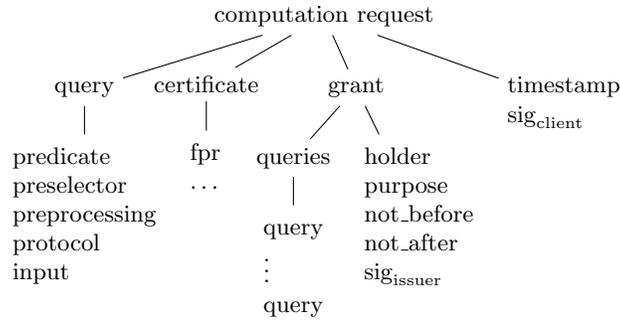
\begin{figure}[t]
\begin{center}
\begin{forest}
  [computation request[query[predicate\\ preselector\\ preprocessing\\ protocol\\ input, align=left]][certificate[fpr\\\ldots, align=left]][grant[queries[query\\\vdots\\query, align=left]][holder\\ purpose\\not\_before\\ not\_after\\$\mathrm{sig}_{\mathrm{issuer}}$, align=left]][timestamp\\$\mathrm{sig}_{\mathrm{client}}$, align=left]]
\end{forest}
\end{center}
\caption{Structure of a computation request}
\label{fig:computation_request}
\end{figure}

\paragraph{Request formats}
There are two request types:
The purpose of the \emph{grant request} from the client is to obtain an authorization grant stating access permission.
A request $r_{c,grant}$ of client $c$ consists of the following components (cf. Figure~\ref{fig:grant_query}):
The \emph{certificate} states the identity of the client.
\iflongversion
This part can be omitted if the certificate is already available on the gateway or made available via the TLS connection providing the request.
\fi
\emph{Queries} contains all queries the clients demands access to.
Each contains the attributes as shown in  Figure~\ref{fig:grant_query}.
${sig}_{client}$ serves authentication and integrity-protection of the request.
The purpose of the second request, the \emph{computation request}, is to actually obtain computed data.
A request $r_{c,comp}$ of client $c$ consists of the following components (cf. Figure~\ref{fig:computation_request}):
The \emph{query} contains the \emph{predicate} and the other characterizing attributes (cf. Section~\ref{sec:query_format}).
The \emph{certificate} is as described in the previous request type.
The \emph{grant} is the authorization grant which states permission to obtain the data in question.
It is the answer of the gateway answer to the previous request type.
The \emph{holder} is the owner identifier of the \emph{grant}.
\emph{not\_before} and \emph{not\_after} specify the time frame of validity.
The \emph{queries} of the \emph{grant} mirror the queries to be allowed for the holder.
\emph{Timestamp} reflects the time when the computation request has been created.
${sig}_{issuer}$ states the permission given by the issuing entity.

\paragraph{Accountability of requests}
Based on the transparency of computation requests which are forwarded to peers, accountability is achieved by persisting the request data structure.
This is extended by a signature of the accepting gateway and optionally the result of the computation\iflongversion{} (cf. Figure~\ref{fig:accountable_entry})\fi.
\iflongversion
\begin{figure}
\begin{center}
\begin{forest}
  [entry[accepted request[computation request[\ldots]][$\mathrm{sig}_{\mathrm{gateway}}$]][result[value][$\mathrm{sig}_{\mathrm{peer}}$]]]
\end{forest}
\end{center}
\caption{Structure of entry for accountability logs}
\label{fig:accountable_entry}
\end{figure}
\fi

\subsection{Protocols}
\label{sec:protocols}
We dissect the client/gateway interaction into three independent protocols:
\inlineEnum{requestTypes}\emph{Metadata request},
\inlineEnum{requestTypes}\emph{grant request},
\inlineEnum{requestTypes}\emph{computation request}.

Here, we model the state of the gateway to be $(\queries, \policies)$ where $\queries$ is the finite set of queries made available by the gateway.
Each query has the structure as shown in Figure~\ref{fig:computation_request}.
$\policies : (query, client, context) \rightarrow \{\mathit{true}, \mathit{false}\}$ is an access control structure; it takes a tuple of a query, a requesting client and a context and returns whether or not access is permitted.

\paragraph{Metadata request}
This is the first interaction between the client and the gateway that requests meta information about the data being available via the gateway.
The gateway answers with a list of all $q \in \queries$.

\begin{figure}
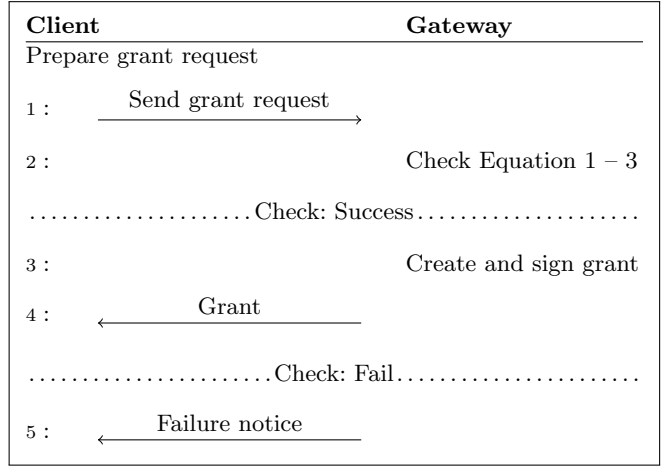

\fbox{%
\pseudocode{%
 \textbf{Client} \< \textbf{Gateway} \\[][\hline]
 \text{Prepare grant request}  \\
\pcln \sendmessageright{top=Send grant request} \< \< \< \\
\pcln \< \text{Check Equation~\ref{eqn:grant_cert_check} --  \ref{eqn:grant_phi_check}} \pclb
 \pcintertext[dotted]{Check: Success}
\pcln  \< \text{Create and sign grant} \\
\pcln  \sendmessageleft{top=Grant} \pclb
 \pcintertext[dotted]{Check: Fail}
\pcln  \sendmessageleft{top=Failure notice}
 }
}
\caption{Grant Request Protocol. In a grant request clients demand an authorization to execute a specified set of queries. The gateway checks, whether or not the requesting client is allowed to access the corresponding data. If yes, an authorization grant is created and handed over to the client.}
\label{fig:grant_protocol}
\end{figure}

\paragraph{Grant request}
Afterwards, clients can demand access to certain information by requesting a corresponding grant (cf. Figure~\ref{fig:grant_protocol}):
They send a grant request $r_{c,grant}$ to the gateway where the checks described below are performed.
If they are successful, the gateway creates and signs a corresponding authorization grant.
This is sent back to the client.

\begin{figure}
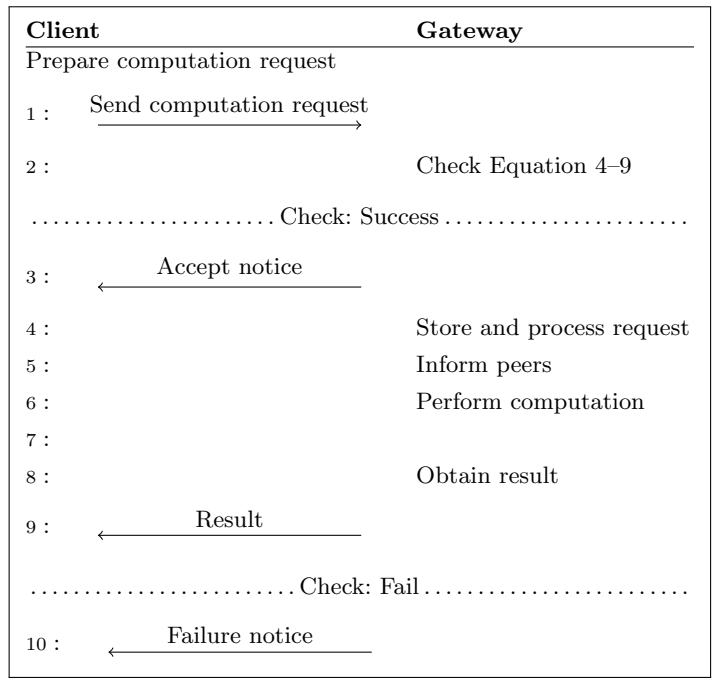

\fbox{%
\pseudocode{%
 \textbf{Client} \< \textbf{Gateway} \\[][\hline]
 \text{Prepare computation request}  \\
\pcln \sendmessageright{top=Send computation request} \< \< \< \\
 \pcln \< \text{Check Equation~\ref{eqn:fpr}--\ref{eqn:query_inclusion}} \pclb
 \pcintertext[dotted]{Check: Success}
\pcln  \sendmessageleft{top=Accept notice} \< \< \< \\
\pcln \label{prot:store}  \< \text{Store and process request} \< \< \\
\pcln \label{prot:inform} \< \text{Inform peers} \< \< \\
\pcln \label{prot:perform} \< \text{Perform computation} \< \< \\
\pcln \< \text{} \\
\pcln \label{prot:obtain} \< \text{Obtain result}  \< \< \\
\pcln \label{prot:result}   \sendmessageleft{top=Result} \< \< \<  \pclb
  \pcintertext[dotted]{Check: Fail}
\pcln   \sendmessageleft{top=Failure notice} \< \< \<
  }
 }
\caption{Computation Request Protocol. A client demands for the results of a computation specified in its request. If the request is valid, the client is notified about acceptance and processing of the request is performed in lines \ref{prot:store} -- \ref{prot:perform} as described in \cite{vonMaltitz2018b}. The result, coming back in line \ref{prot:obtain} is then forwarded to the client in line \ref{prot:result}.}
\label{fig:comp_protocol}
\end{figure}

\textbf{Verification}
Given a request $r_{c,grant}$ of client $c$, the validity of the client certificate (Equation~\ref{eqn:grant_cert_check}) is checked and whether the requesting client possesses the corresponding private key.\footnote{When using TLS, this is already handled during TLS session establishment.}
Similarly, validity of the request signature (Equation~\ref{eqn:grant_sig_check}) is verified.
Then the semantics of the queries are checked (Equation~\ref{eqn:grant_phi_check}) using $\Phi$.
The parameters are set as follows: The \emph{query} reflects the demanded data in a form as described above.
The \emph{client} is represented by its certificate.
The \emph{context} is the current state of the gateway, this encompasses the information about currently connected peers and environment information like the current time.
\begin{align}
  \label{eqn:grant_cert_check}
  \mathit{verify}(r_{c,grant}.\mathit{cert})\\
  \label{eqn:grant_sig_check}
  \mathit{verify}(r_{c,grant}.sig_{\mathit{client}}, r_{c,grant}.\mathit{cert})\\
  \label{eqn:grant_phi_check}
  \forall q \in r_{c,grant}.queries: \Phi(q, r_{c,grant}.\mathit{cert}, context)
\end{align}
\paragraph{Computation request}
Requests of this type are performed repeatedly during productive use in order to obtain aggregated data from the peers.
The protocol is shown in Figure~\ref{fig:comp_protocol}.
The client sends a computation request (cf. Figure~\ref{fig:computation_request}) to the gateway.
If the checks as described below are successful, the request is accepted and transformed into an SMC session.
The result of the SMC session is the requested information.
It is signed by the peers and, since the certificate of the client is available to the peers, encrypted for the client.
The encrypted result can then be forwarded by the gateway to the requesting client.

\textbf{Verification}
The gateway and the peers play different roles regarding access control and intervenability; hence they validate different aspects of the request.

Given a request $r_{c,comp}$ of client $c$,
the gateway first checks whether the holder matches the requesting client and the client certificate is valid (Equation~\ref{eqn:fpr}),
it checks the authenticity of the request (Equation~\ref{eqn:req_auth}),
then it verifies whether the authorization grant itself is valid (Equation~\ref{eqn:timea} -- \ref{eqn:issuer}).
\begin{align}
  \label{eqn:fpr}
  \mathit{r_{c,comp}.grant.holder} = \mathit{c.cert.fpr} \wedge \mathit{verify}(c.\mathit{cert})\\
  \label{eqn:req_auth}
  \mathit{verify}(sig_{\mathit{client}}, c.\mathit{cert})\\
  \label{eqn:timea}
  {\mathit{r_{c,comp}.grant.not\_before}} \leq now \\
  \label{eqn:timeb}
  now \leq {\mathit{r_{c,comp}.grant.not\_after}}  \\
  \label{eqn:issuer}
  \mathit{verify}(\mathit{issuer.cert}) \wedge \mathit{verify}(sig_{\mathit{issuer}}, \mathit{issuer.cert})
\end{align}
After checking formal validity, the validity of the request itself is verified, i.e., whether the grant supports the stated query.
This is \iflongversion either \fi realized by checking for the
inclusion of the query in the permitted set (Equation~\ref{eqn:query_inclusion})%
\iflongversion%
, or,
if the issuer is able to create more general predicates than singular requests,
satisfaction of the request query by one of the grant queries (Equation~\ref{eqn:query_satisfaction}).
\else%
.
\fi
\begin{align}
  \label{eqn:query_inclusion}
  \exists q \in \mathit{r_{c,comp}.grant.queries}: q = r_{c,comp}.\mathit{query}%
\iflongversion%
\\
  \label{eqn:query_satisfaction}
  \exists q \in \mathit{r_{c,comp}.grant.queries}: q \vdash r_{c,comp}.\mathit{query}
\fi
\end{align}
Verification by the peers happens when the session is communicated to them (Step \ref{prot:inform} in Figure~\ref{fig:comp_protocol}).
Besides rechecking abovementioned checks,
each peer $p \in P$ can have a set $\policies_p$ of local policies which defines how their data may be used and the corresponding privacy constraints.
Satisfaction of these policies can also be checked (Equation~\ref{eqn:ppolicy}).
This e.g. can include a check for recency of the computation request in order to prevent replay attacks.

\begin{align}
  \label{eqn:ppolicy}
  \policies_p \vdash (c, r_{c,comp}.query)
\end{align}

\section{Discussion}
\label{sec:discussion}
In the previous sections we presented  an approach to fulfill the requirements stated in Section~\ref{sec:goal_statement}.
The computation request and the corresponding protocol allow third parties to query for data computed by SMC (\ref{req:queries}).
By doing so, securely computed information becomes accessible to outside clients.
Access on the computed results can be controlled by authorization grants required for computation requests (\ref{req:accesscontrol}).
This enables controlled data flow when serving a heterogeneous set of deployed services.

Making the request verification independent from the access control structure $\Phi$ and its \emph{context} parameter, i.e. not relying on complex state of the gateway for access control but on transferable documents, allows to use the same data structures to fulfill desired privacy-protection goals:
Peers are informed about upcoming computations and their context if they are involved, since
the forwarded requests include the query and the corresponding authorization grant of the requesting client (\ref{req:transparency}).
An authenticated history of data access and usage can be built (\ref{req:accountability}), since authenticated requests can also be persisted by each peer.
The signatures of the client, the gateway and the peer ensure that integrity of request and corresponding grant is protected.
Giving peers the ability to verify this request using their own local policies $\Phi_p$  and allowing them to veto against requested computations enables intervenability (\ref{req:intervenability}).
\iflongversion
This is based on the fundamental necessity of SMC that computation cannot happen without cooperation of the data possessing peers.
\fi
As a consequence, data minimization is supported:
Data sources can to make sure that there are not more or different computations performed than they expect to happen.

Regarding possible attacks from Section~\ref{sec:sec_model}, our solution performs as follows:
\ref{att:peercurious}, \ref{att:peercorrectness}, and \ref{att:clientcorrelate} have been excluded by design.
A gateway cannot forge computation requests (\ref{att:gatewayrequest}), since it cannot provide a valid signature for Equation~\ref{eqn:req_auth}.
Replay attacks using former computation requests is prevented by adding a timestamp to the request.
The gateway cannot collect data requested by valid clients (\ref{att:gatewaycurious}), since the obtained result (cf. Figure~\ref{fig:comp_protocol}, step \ref{prot:obtain}) is encrypted for the receiving client.
Similarly, the result cannot be changed by the gateway (\ref{att:gatewaytamper}), since the result is signed by the peers.
Prevention of unauthorized access by clients (\ref{att:clientaccess}) is a fundamental goal of our work and achieved
by providing grants which state the permissions of every individual client.
\iflongversion
This allows gateway and peers to verify whether and why a given computation request should be allowed for a certain client.
\fi
Our protocol introduces no new leaks for the information computed by SMC.
Hence, the security properties of the utilized SMC realization apply:
The state of the art~\cite{BristolMPC,BreakingSPDZ,MASCOT,Keller2018} is already secure against $n-1$ maliciously colluding peers.
Since the data owners' device always participates in the computation when using their data, security of the own raw data is already achieved by guaranteeing that
the own device is not compromized.

\section{Evaluation}
\label{sec:evaluation}
For evaluation of our approach we implemented a prototype in python and performed measurements
of the protocols presented (cf. Figure~\ref{fig:grant_protocol} and \ref{fig:comp_protocol}).
\subsection{Setup}
\paragraph{Scenario}
As scenario we consider  a single floor in a smart building.
This implies around 10 to 30 peers which are connected to a single gateway.
The gateway is assumed to be decent commodity hardware\iflongversion{} with several CPUs and a sufficient amount of main memory\fi.
The network is an intranet with low latency, a typical throughput and no packet loss.

\paragraph{Hardware and System}
We used 4 hosts in our setup.
These are equipped with
\iflongversion
Intel Xeon E3-1265L V2  CPU with
\fi
eight cores at 2.50\,GHz \iflongversion{} and a cache size of 8192\,KB \fi and a main memory of
15.780\,MB.
They have 1\,Gbit networking interfaces and
are arranged in a star topology\iflongversion, all hosts are connected via a single switch\fi.
The default link latency is around 0.18\,ms.
\iflongversion

The hosts use a dedicated PXE server to boot from.
This server provides an image of Debian Stretch (9.4) using a 4.9.0 Linux kernel.
\else
As operating system we use Debian Stretch (9.4) based on a 4.9.0 Linux kernel.
\fi

The roles of the gateway, the client and the peers have to be reflected in the setup.
We deployed the gateway and the client on individual hosts.
Furthermore, a single peer was deployed on a dedicated host; all other peers were started as processes on a single further machine.

\paragraph{Implementation}
The prototype is implemented in python as a flask~1.0.2 application
and executed using python~3.5.3.
Data is stored in a mongodb version 3.2.11 and as authorization backend authzforce 8.0.1 \cite{authzforce}
is used which is located on the gateway host\iflongversion{} and connected via a REST interface\fi.
The flask application is served by uwsgi~2.0.17.1.
Uwsgi is executed with a single process and eight threads, if not told otherwise.
The queue for unanswered requests has a limit of 100 entries.
\iflongversion
If the queue is full, further requests are immediately rejected.

During computation request, the gateway has to contact peers in order to allow them to verify the client request (Step \ref{prot:inform} in Figure~\ref{fig:comp_protocol}).
For doing so, the peer component also features a REST interface which is made available by uwsgi.
\fi
When contacting the peers, the gateway spawns a thread for each peer in order to allow simultaneous waiting for all responses.

For testing purposes no actual SMC component was connected to our querying framework.
This allows to measure the overhead of our components without depending on the performance
characteristics of a chosen SMC implementation.
We decided for this omission, since we assessed the performance of a state of the art SMC implementation elsewhere \cite{vonMaltitz2018a}.

\paragraph{Method}
\iflongversion
The client has been scripted to offer identical requests with different frequencies.
For each measured frequency step, the client offered the desired load for either 30s (grant request protocol) or 60s (computation request protocol).
\fi
We measured the duration of client requests handled by the gateway (\emph{latency}).
The measured time begins when the request is handed to our custom code; it ends when the final response is handed back to uwsgi.
In the results, we show the median, the 0.25 and the 0.75 quantile.
For each frequency, we captured the amount of requests the gateway was able to handle successfully per second (\emph{throughput}).
Lastly we recorded the state of the request queue.
\iflongversion
Uwsgi allows to query its current degree of saturation via a http stats interface.
The queue state was always queried immediately after the last request was sent, ensuring that the complete load has reached the gateway but was not yet handled by it.
\fi

\newlength\figureheight
\newlength\figurewidth
\setlength\figureheight{5cm}
\setlength\figurewidth{\columnwidth}

\subsection{Results}

\paragraph{Grant Request Protocol}
The grant request protocol is carried out when a client aims to obtain further access rights to data queries offered by the gateway.
\iflongversion
The frequency of this request depends on the volatility of the environment:
In a setting where peers and clients are deployed a single time at the beginning of system lifetime and further changes are seldom, requests are performed a small number of times per client.
Every change might make further grant requests necessary.
In a setting where clients change often or data querying is based on individual user interaction instead of automated predetermined processes, frequency can increase to small number of requests per second during peak times.
Furthermore, renewal of grants makes repetition of requests necessary.
\else
This only happens when new clients are deployed or permissions change over time.
We assume that a small number of requests per second is only exceeded during peak times.
\fi

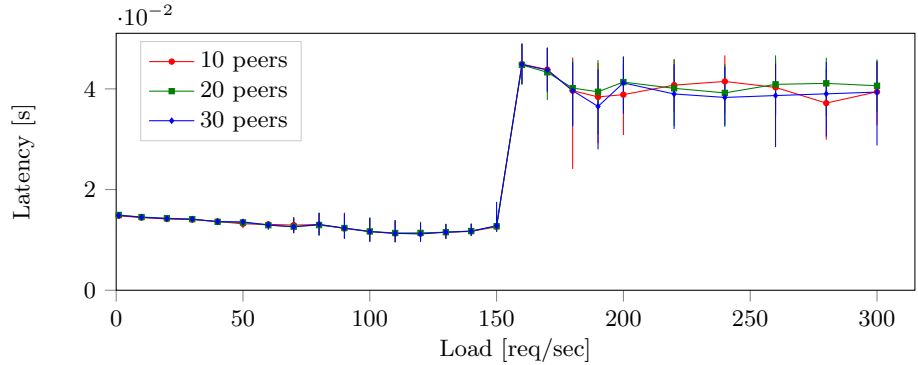
\begin{figure}[t]
\begin{tikzpicture}

\begin{axis}[
xlabel={Load [req/sec]},
ylabel={Latency [s]},
xmin=0, xmax=314.95,
ymin=0, ymax=0.05103756781125,
width=\figurewidth,
height=\figureheight,
tick align=outside,
tick pos=left,
x grid style={white!69.01960784313725!black},
y grid style={white!69.01960784313725!black},
legend entries={{10 peers},{20 peers},{30 peers}},
legend style={at={(0.03,0.97)}, anchor=north west, draw=white!80.0!black},
legend cell align={left}
]
\path [draw=red] (axis cs:1,0.01446014645)
--(axis cs:1,0.015024185225);

\path [draw=red] (axis cs:10,0.014137089275)
--(axis cs:10,0.01460039615);

\path [draw=red] (axis cs:20,0.013901054875)
--(axis cs:20,0.014449000325);

\path [draw=red] (axis cs:30,0.01369720695)
--(axis cs:30,0.0143595338);

\path [draw=red] (axis cs:40,0.013074696075)
--(axis cs:40,0.014079690025);

\path [draw=red] (axis cs:50,0.012311518175)
--(axis cs:50,0.01382893325);

\path [draw=red] (axis cs:60,0.012179851575)
--(axis cs:60,0.013726115225);

\path [draw=red] (axis cs:70,0.0114958882)
--(axis cs:70,0.01455765965);

\path [draw=red] (axis cs:80,0.010956525775)
--(axis cs:80,0.015350580225);

\path [draw=red] (axis cs:90,0.01026207205)
--(axis cs:90,0.01542788745);

\path [draw=red] (axis cs:100,0.00965338945)
--(axis cs:100,0.0143767595);

\path [draw=red] (axis cs:110,0.009500861175)
--(axis cs:110,0.0138813853);

\path [draw=red] (axis cs:120,0.0097596049)
--(axis cs:120,0.013406872775);

\path [draw=red] (axis cs:130,0.01018840075)
--(axis cs:130,0.0130776167);

\path [draw=red] (axis cs:140,0.010753095175)
--(axis cs:140,0.01318991185);

\path [draw=red] (axis cs:150,0.011560738075)
--(axis cs:150,0.0153300762);

\path [draw=red] (axis cs:160,0.04104423525)
--(axis cs:160,0.048896133925);

\path [draw=red] (axis cs:170,0.0397336483)
--(axis cs:170,0.04823350905);

\path [draw=red] (axis cs:180,0.0240989327)
--(axis cs:180,0.046239852925);

\path [draw=red] (axis cs:190,0.0291711092)
--(axis cs:190,0.045121789);

\path [draw=red] (axis cs:200,0.0308307409)
--(axis cs:200,0.045045912225);

\path [draw=red] (axis cs:220,0.035422444375)
--(axis cs:220,0.04585397245);

\path [draw=red] (axis cs:240,0.03526991605)
--(axis cs:240,0.0466482639);

\path [draw=red] (axis cs:260,0.0350541472)
--(axis cs:260,0.045667469475);

\path [draw=red] (axis cs:280,0.02990818025)
--(axis cs:280,0.04394459725);

\path [draw=red] (axis cs:300,0.03277421)
--(axis cs:300,0.0454061031);

\path [draw=green!50.0!black] (axis cs:1,0.0146469474)
--(axis cs:1,0.01519441605);

\path [draw=green!50.0!black] (axis cs:10,0.01425218585)
--(axis cs:10,0.01472085715);

\path [draw=green!50.0!black] (axis cs:20,0.014062345075)
--(axis cs:20,0.014541745175);

\path [draw=green!50.0!black] (axis cs:30,0.0137339234)
--(axis cs:30,0.014427185025);

\path [draw=green!50.0!black] (axis cs:40,0.012974679475)
--(axis cs:40,0.014030575725);

\path [draw=green!50.0!black] (axis cs:50,0.012630939475)
--(axis cs:50,0.013966619925);

\path [draw=green!50.0!black] (axis cs:60,0.01210600135)
--(axis cs:60,0.01372337345);

\path [draw=green!50.0!black] (axis cs:70,0.011402726175)
--(axis cs:70,0.014412403125);

\path [draw=green!50.0!black] (axis cs:80,0.01087802645)
--(axis cs:80,0.0153917074);

\path [draw=green!50.0!black] (axis cs:90,0.01024585965)
--(axis cs:90,0.01526266335);

\path [draw=green!50.0!black] (axis cs:100,0.00971621275)
--(axis cs:100,0.01443231105);

\path [draw=green!50.0!black] (axis cs:110,0.009579420075)
--(axis cs:110,0.0139080882);

\path [draw=green!50.0!black] (axis cs:120,0.009808540375)
--(axis cs:120,0.01360923055);

\path [draw=green!50.0!black] (axis cs:130,0.010226070875)
--(axis cs:130,0.0131700039);

\path [draw=green!50.0!black] (axis cs:140,0.010794579975)
--(axis cs:140,0.013265967375);

\path [draw=green!50.0!black] (axis cs:150,0.011609673475)
--(axis cs:150,0.015863716575);

\path [draw=green!50.0!black] (axis cs:160,0.040807187575)
--(axis cs:160,0.048883557325);

\path [draw=green!50.0!black] (axis cs:170,0.0377987623)
--(axis cs:170,0.04818880555);

\path [draw=green!50.0!black] (axis cs:180,0.0329093933)
--(axis cs:180,0.0457291603);

\path [draw=green!50.0!black] (axis cs:190,0.03098750115)
--(axis cs:190,0.045719862);

\path [draw=green!50.0!black] (axis cs:200,0.0357992649)
--(axis cs:200,0.0462653637);

\path [draw=green!50.0!black] (axis cs:220,0.032626152)
--(axis cs:220,0.0458946228);

\path [draw=green!50.0!black] (axis cs:240,0.032419741175)
--(axis cs:240,0.0449209809);

\path [draw=green!50.0!black] (axis cs:260,0.03455877305)
--(axis cs:260,0.04664468765);

\path [draw=green!50.0!black] (axis cs:280,0.03471207615)
--(axis cs:280,0.046180844325);

\path [draw=green!50.0!black] (axis cs:300,0.0353739262)
--(axis cs:300,0.04585832355);

\path [draw=blue] (axis cs:1,0.0145154595)
--(axis cs:1,0.0150909424);

\path [draw=blue] (axis cs:10,0.014234721675)
--(axis cs:10,0.0146996975);

\path [draw=blue] (axis cs:20,0.014005959)
--(axis cs:20,0.0145646334);

\path [draw=blue] (axis cs:30,0.01370227335)
--(axis cs:30,0.0144481659);

\path [draw=blue] (axis cs:40,0.013051629075)
--(axis cs:40,0.01408332585);

\path [draw=blue] (axis cs:50,0.012829601775)
--(axis cs:50,0.0140299201);

\path [draw=blue] (axis cs:60,0.0120487213)
--(axis cs:60,0.013689100775);

\path [draw=blue] (axis cs:70,0.011311113825)
--(axis cs:70,0.014400780175);

\path [draw=blue] (axis cs:80,0.01088190075)
--(axis cs:80,0.015390634575);

\path [draw=blue] (axis cs:90,0.010264396675)
--(axis cs:90,0.0152094364);

\path [draw=blue] (axis cs:100,0.0096330047)
--(axis cs:100,0.01439219715);

\path [draw=blue] (axis cs:110,0.009562194375)
--(axis cs:110,0.013904273525);

\path [draw=blue] (axis cs:120,0.009598374375)
--(axis cs:120,0.013368189375);

\path [draw=blue] (axis cs:130,0.010273277775)
--(axis cs:130,0.0131868124);

\path [draw=blue] (axis cs:140,0.010827004875)
--(axis cs:140,0.0131934285);

\path [draw=blue] (axis cs:150,0.011674404175)
--(axis cs:150,0.017542779425);

\path [draw=blue] (axis cs:160,0.04094272855)
--(axis cs:160,0.0490596294);

\path [draw=blue] (axis cs:170,0.0393686891)
--(axis cs:170,0.04809701445);

\path [draw=blue] (axis cs:180,0.0325119495)
--(axis cs:180,0.0452382565);

\path [draw=blue] (axis cs:190,0.02799582485)
--(axis cs:190,0.04384434225);

\path [draw=blue] (axis cs:200,0.03507018085)
--(axis cs:200,0.04645955565);

\path [draw=blue] (axis cs:220,0.0320501328)
--(axis cs:220,0.0448980331);

\path [draw=blue] (axis cs:240,0.032756030575)
--(axis cs:240,0.04442340135);

\path [draw=blue] (axis cs:260,0.0284464359)
--(axis cs:260,0.0450364351);

\path [draw=blue] (axis cs:280,0.03053826095)
--(axis cs:280,0.0453463793);

\path [draw=blue] (axis cs:300,0.028755307225)
--(axis cs:300,0.0455252528);

\addplot [red, mark=*, mark size=1, mark options={solid}]
table {%
1 0.014783144
10 0.01443588735
20 0.01415228845
30 0.0140503645
40 0.0136567354
50 0.0131924152
60 0.0130883455
70 0.0129388571
80 0.0130610466
90 0.0122680664
100 0.0116764307
110 0.01128315925
120 0.01127898695
130 0.0115154982
140 0.01169729235
150 0.0126159191
160 0.0449155569
170 0.0438241959
180 0.03963792325
190 0.0384044647
200 0.0388534069
220 0.0407449007
240 0.04147541525
260 0.0402884483
280 0.0371732712
300 0.0394170284
};
\addplot [green!50.0!black, mark=square*, mark size=1, mark options={solid}]
table {%
1 0.0149562359
10 0.0145406723
20 0.01430499555
30 0.01413011555
40 0.0135933161
50 0.01346838475
60 0.01292312145
70 0.01263809205
80 0.01294004915
90 0.0123615265
100 0.01168346405
110 0.011377573
120 0.0113954544
130 0.01152825355
140 0.01176166535
150 0.01271402835
160 0.04475617405
170 0.0432920456
180 0.040158987
190 0.0394260883
200 0.0413513184
220 0.0401077271
240 0.0392012596
260 0.0408966541
280 0.04109358785
300 0.04062092305
};
\addplot [blue, mark=diamond*, mark size=1, mark options={solid}]
table {%
1 0.0148832798
10 0.01450395585
20 0.0142431259
30 0.01410865785
40 0.01364326475
50 0.0135781765
60 0.012966156
70 0.01258969305
80 0.0130814314
90 0.01234066485
100 0.011646986
110 0.01130259035
120 0.0111820698
130 0.01154363155
140 0.0117304325
150 0.0128194094
160 0.04482483865
170 0.0437538624
180 0.0396482944
190 0.0365114212
200 0.0411863327
220 0.0389873981
240 0.03830206395
260 0.038672924
280 0.03901290895
300 0.0393614769
};
\end{axis}

\end{tikzpicture}
\vspace{-6mm}
  \caption{Grant Request Protocol: The duration of handling a single request inside the gateway component depending on the amount of requests performed by the client. }
  \label{fig:ag_gateway_load_coarse}
\end{figure}

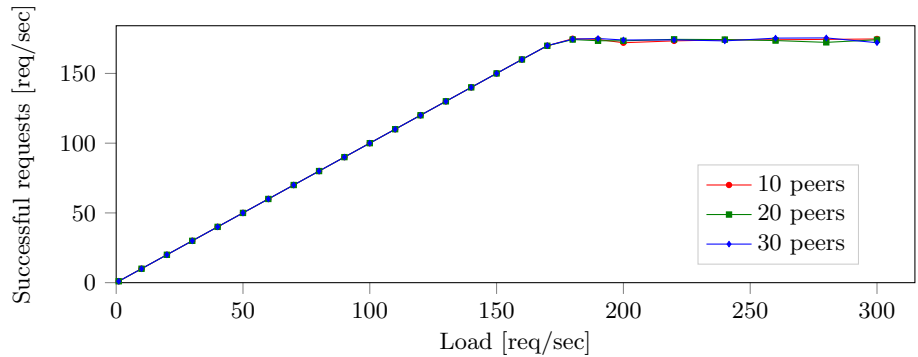
\begin{figure}[t]
\begin{tikzpicture}

\begin{axis}[
xlabel={Load [req/sec]},
ylabel={Successful requests [req/sec]},
xmin=0, xmax=314.95,
ymin=0, ymax=184.19,
width=\figurewidth,
height=\figureheight,
tick align=outside,
tick pos=left,
x grid style={white!69.01960784313725!black},
y grid style={white!69.01960784313725!black},
legend entries={{10 peers},{20 peers},{30 peers}},
legend style={at={(0.93,0.07)}, anchor=south east, draw=white!80.0!black},
legend cell align={left}
]
\addplot [red, mark=*, mark size=1, mark options={solid}]
table {%
1 1
10 10
20 20
30 30
40 40
50 50
60 60
70 70
80 80
90 90
100 100
110 110
120 120
130 130
140 140
150 150
160 160
170 169.833333333333
180 174.866666666667
190 174.366666666667
200 172
220 173.333333333333
240 174.066666666667
260 174.133333333333
280 174.4
300 174.7
};
\addplot [green!50.0!black, mark=square*, mark size=1, mark options={solid}]
table {%
1 1
10 10
20 20
30 30
40 40
50 50
60 60
70 70
80 80
90 90
100 100
110 110
120 120
130 130
140 140
150 150
160 160
170 169.833333333333
180 174.3
190 173.433333333333
200 173.366666666667
220 174.433333333333
240 174.266666666667
260 173.566666666667
280 172.266666666667
300 174.133333333333
};
\addplot [blue, mark=diamond*, mark size=1, mark options={solid}]
table {%
1 1
10 10
20 20
30 30
40 40
50 50
60 60
70 70
80 80
90 90
100 100
110 110
120 120
130 130
140 140
150 150
160 160
170 169.933333333333
180 174.566666666667
190 175.033333333333
200 173.966666666667
220 174.033333333333
240 173.333333333333
260 175.3
280 175.466666666667
300 172.033333333333
};
\end{axis}

\end{tikzpicture}
\vspace{-6mm}
  \caption{Grant Request Protocol: The amount of successfully answered requests depending on the amount of requests performed by the client. Up to a load of 170 requests/s, throughput increases proportionally and no drops occur. Afterwards, the queue is filled and throughput stagnates on this level.}
  \label{fig:ag_gateway_throughput}
\end{figure}

This amount of requests is well supported.
Even under a load of $\ge 100$ requests per second, answer time stays below 20\,ms (Figure~\ref{fig:ag_gateway_load_coarse}).
The queue of the gateway becomes saturated only after 170 requests/s\iflongversion{} (Figure~\ref{fig:ag_gateway_queue})\fi{}.
Correspondingly, the throughput stagnates at the
\iflongversion
level of 170 requests/s (Figure~\ref{fig:ag_gateway_throughput}); all remaining requests are dropped.
\else
same point (Figure~\ref{fig:ag_gateway_throughput}).
\fi
Since no peer interaction happens, performance is independent of their number.

\iflongversion
\begin{figure}[t]
\begin{tikzpicture}

\begin{axis}[
xlabel={Load [req/sec]},
ylabel={Queue length [\#]},
xmin=0, xmax=314.95,
ymin=0, ymax=106.05,
width=\figurewidth,
height=\figureheight,
tick align=outside,
tick pos=left,
x grid style={white!69.01960784313725!black},
y grid style={white!69.01960784313725!black},
legend entries={{10 peers},{20 peers},{30 peers}},
legend style={at={(0.03,0.97)}, anchor=north west, draw=white!80.0!black},
legend cell align={left}
]
\addplot [red, mark=*, mark size=1, mark options={solid}]
table {%
1 0
10 0
20 0
30 0
40 0
50 0
60 0
70 0
80 0
90 0
100 0
110 0
120 0
130 0
140 0
150 1
160 101
170 100
180 100
190 100
200 101
220 99
240 101
260 101
280 100
300 100
};
\addplot [green!50.0!black, mark=square*, mark size=1, mark options={solid}]
table {%
1 0
10 0
20 0
30 0
40 0
50 0
60 0
70 0
80 0
90 0
100 0
110 0
120 0
130 0
140 0
150 0
160 44
170 100
180 100
190 100
200 101
220 100
240 101
260 101
280 100
300 101
};
\addplot [blue, mark=diamond*, mark size=1, mark options={solid}]
table {%
1 0
10 0
20 0
30 0
40 0
50 0
60 0
70 0
80 0
90 0
100 0
110 0
120 0
130 0
140 0
150 3
160 53
170 100
180 99
190 99
200 101
220 99
240 101
260 101
280 101
300 96
};
\end{axis}

\end{tikzpicture}
\vspace{-6mm}
  \caption{Grant Request Protocol: Saturation of the uwsgi request queue depending on the amount of requests performed by the client. As long as the queue is not saturated, all requests can be answered successfully. During higher loads, requests are dropped during connection attempt. }
  \label{fig:ag_gateway_queue}
\end{figure}
\fi

\paragraph{Computation Request Protocol}
The computation request protocol is always carried out when a computation on actual sensor  data is queried.
With polling every second per client, and multiple clients being connected, multiple requests per second can be expected.

%

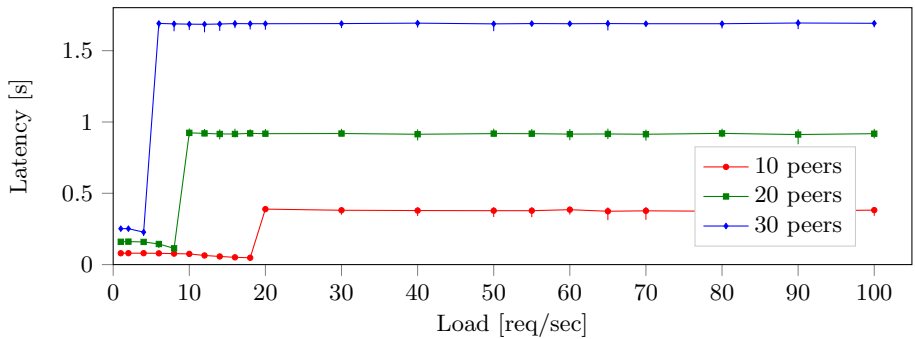
\begin{figure}[t]
\begin{tikzpicture}

\begin{axis}[
xlabel={Load [req/sec]},
ylabel={Latency [s]},
xmin=0, xmax=104.95,
ymin=0, ymax=1.8016,
width=\figurewidth,
height=\figureheight,
tick align=outside,
tick pos=left,
x grid style={white!69.01960784313725!black},
y grid style={white!69.01960784313725!black},
legend entries={{10 peers},{20 peers},{30 peers}},
legend style={at={(0.93,0.07)}, anchor=south east, draw=white!80.0!black},
legend cell align={left}
]
\path [draw=red] (axis cs:1,0.078)
--(axis cs:1,0.081);

\path [draw=red] (axis cs:2,0.078)
--(axis cs:2,0.081);

\path [draw=red] (axis cs:4,0.078)
--(axis cs:4,0.081);

\path [draw=red] (axis cs:6,0.078)
--(axis cs:6,0.081);

\path [draw=red] (axis cs:8,0.076)
--(axis cs:8,0.079);

\path [draw=red] (axis cs:10,0.071)
--(axis cs:10,0.077);

\path [draw=red] (axis cs:12,0.06)
--(axis cs:12,0.072);

\path [draw=red] (axis cs:14,0.052)
--(axis cs:14,0.067);

\path [draw=red] (axis cs:16,0.048)
--(axis cs:16,0.056);

\path [draw=red] (axis cs:18,0.046)
--(axis cs:18,0.05);

\path [draw=red] (axis cs:20,0.369)
--(axis cs:20,0.408);

\path [draw=red] (axis cs:30,0.35)
--(axis cs:30,0.404);

\path [draw=red] (axis cs:40,0.341)
--(axis cs:40,0.40225);

\path [draw=red] (axis cs:50,0.334)
--(axis cs:50,0.403);

\path [draw=red] (axis cs:55,0.333)
--(axis cs:55,0.402);

\path [draw=red] (axis cs:60,0.351)
--(axis cs:60,0.407);

\path [draw=red] (axis cs:65,0.31175)
--(axis cs:65,0.401);

\path [draw=red] (axis cs:70,0.314)
--(axis cs:70,0.404);

\path [draw=red] (axis cs:80,0.3)
--(axis cs:80,0.402);

\path [draw=red] (axis cs:90,0.316)
--(axis cs:90,0.402);

\path [draw=red] (axis cs:100,0.342)
--(axis cs:100,0.404);

\path [draw=green!50.0!black] (axis cs:1,0.15875)
--(axis cs:1,0.162);

\path [draw=green!50.0!black] (axis cs:2,0.16)
--(axis cs:2,0.164);

\path [draw=green!50.0!black] (axis cs:4,0.156)
--(axis cs:4,0.161);

\path [draw=green!50.0!black] (axis cs:6,0.116)
--(axis cs:6,0.157);

\path [draw=green!50.0!black] (axis cs:8,0.111)
--(axis cs:8,0.121);

\path [draw=green!50.0!black] (axis cs:10,0.89775)
--(axis cs:10,0.954);

\path [draw=green!50.0!black] (axis cs:12,0.9)
--(axis cs:12,0.951);

\path [draw=green!50.0!black] (axis cs:14,0.878)
--(axis cs:14,0.948);

\path [draw=green!50.0!black] (axis cs:16,0.893)
--(axis cs:16,0.9525);

\path [draw=green!50.0!black] (axis cs:18,0.898)
--(axis cs:18,0.949);

\path [draw=green!50.0!black] (axis cs:20,0.898)
--(axis cs:20,0.951);

\path [draw=green!50.0!black] (axis cs:30,0.9)
--(axis cs:30,0.952);

\path [draw=green!50.0!black] (axis cs:40,0.8705)
--(axis cs:40,0.948);

\path [draw=green!50.0!black] (axis cs:50,0.90025)
--(axis cs:50,0.953);

\path [draw=green!50.0!black] (axis cs:55,0.899)
--(axis cs:55,0.95);

\path [draw=green!50.0!black] (axis cs:60,0.872)
--(axis cs:60,0.95);

\path [draw=green!50.0!black] (axis cs:65,0.88)
--(axis cs:65,0.953);

\path [draw=green!50.0!black] (axis cs:70,0.869)
--(axis cs:70,0.946);

\path [draw=green!50.0!black] (axis cs:80,0.893)
--(axis cs:80,0.953);

\path [draw=green!50.0!black] (axis cs:90,0.844)
--(axis cs:90,0.949);

\path [draw=green!50.0!black] (axis cs:100,0.885)
--(axis cs:100,0.952);

\path [draw=blue] (axis cs:1,0.249)
--(axis cs:1,0.254);

\path [draw=blue] (axis cs:2,0.249)
--(axis cs:2,0.254);

\path [draw=blue] (axis cs:4,0.2)
--(axis cs:4,0.24725);

\path [draw=blue] (axis cs:6,1.65775)
--(axis cs:6,1.70425);

\path [draw=blue] (axis cs:8,1.63625)
--(axis cs:8,1.708);

\path [draw=blue] (axis cs:10,1.6455)
--(axis cs:10,1.698);

\path [draw=blue] (axis cs:12,1.629)
--(axis cs:12,1.702);

\path [draw=blue] (axis cs:14,1.63875)
--(axis cs:14,1.697);

\path [draw=blue] (axis cs:16,1.658)
--(axis cs:16,1.712);

\path [draw=blue] (axis cs:18,1.649)
--(axis cs:18,1.703);

\path [draw=blue] (axis cs:20,1.647)
--(axis cs:20,1.701);

\path [draw=blue] (axis cs:30,1.65875)
--(axis cs:30,1.706);

\path [draw=blue] (axis cs:40,1.662)
--(axis cs:40,1.707);

\path [draw=blue] (axis cs:50,1.637)
--(axis cs:50,1.701);

\path [draw=blue] (axis cs:55,1.6705)
--(axis cs:55,1.701);

\path [draw=blue] (axis cs:60,1.676)
--(axis cs:60,1.698);

\path [draw=blue] (axis cs:65,1.642)
--(axis cs:65,1.714);

\path [draw=blue] (axis cs:70,1.66975)
--(axis cs:70,1.703);

\path [draw=blue] (axis cs:80,1.656)
--(axis cs:80,1.702);

\path [draw=blue] (axis cs:90,1.651)
--(axis cs:90,1.718);

\path [draw=blue] (axis cs:100,1.664)
--(axis cs:100,1.705);

\addplot [red, mark=*, mark size=1, mark options={solid}]
table {%
1 0.0795
2 0.08
4 0.08
6 0.079
8 0.077
10 0.075
12 0.064
14 0.057
16 0.051
18 0.048
20 0.389
30 0.381
40 0.379
50 0.378
55 0.378
60 0.385
65 0.374
70 0.377
80 0.374
90 0.376
100 0.382
};
\addplot [green!50.0!black, mark=square*, mark size=1, mark options={solid}]
table {%
1 0.16
2 0.161
4 0.159
6 0.1445
8 0.114
10 0.924
12 0.92
14 0.916
16 0.916
18 0.92
20 0.918
30 0.919
40 0.914
50 0.9185
55 0.918
60 0.915
65 0.916
70 0.914
80 0.92
90 0.912
100 0.918
};
\addplot [blue, mark=diamond*, mark size=1, mark options={solid}]
table {%
1 0.252
2 0.252
4 0.2275
6 1.691
8 1.688
10 1.686
12 1.685
14 1.687
16 1.69
18 1.689
20 1.689
30 1.69
40 1.693
50 1.688
55 1.69
60 1.689
65 1.69
70 1.689
80 1.689
90 1.694
100 1.692
};
\end{axis}

\end{tikzpicture}
\vspace{-6mm}
  \caption{Computation Request Protocol: The duration of handling a single request inside the gateway component depending on the amount of requests performed by the client. This includes forwarding the request to all concerned peers and waiting for their request acceptance. Waiting for the responses of the peers constitutes the biggest part of the overall duration.}
  \label{fig:ir_gateway_load_coarse}
\end{figure}

With 30 peers connected, a single request per second yields a latency of \texttildelow 250ms.
With increasing load this converges to \texttildelow 1.7 seconds per request (Figure~\ref{fig:ir_gateway_load_coarse}).
Varying the number of peers show that each added peer approximately contributes further 50\,ms to the overall latency.
The reason is computational overhead per connection -- mainly signing outgoing messages and verifying the signatures of incoming messages -- which cannot be handled in parallel due to the global interpreter lock in python.
We can propose several mitigations for this overhead:
\inlineEnum{miti} Firstly, we assume a non-prototypical implementation in a language which optimizes more for speed will provide a lower overhead.
Especially using a language which provides real parallelization and/or more efficient implementations of the corresponding actions should decrease this overhead.
\inlineEnum{miti} Caching can be considered: Peers' agreement to computations could be extended by an interval of validity, implicitly allowing later consecutive identical requests without further overhead.
\inlineEnum{miti} Our architecture can be applied hierarchically.
allowing several gateways per area which segment peers into semantically sensible groups, being large enough to ensure unlinkability.
Recursively, gateways can in turn function as higher-level peers for further aggregation.

\iflongversion
\begin{figure}[t]
\begin{tikzpicture}

\begin{axis}[
xlabel={Load [req/sec]},
ylabel={Queue length [\#]},
xmin=0, xmax=104.95,
ymin=0, ymax=108.15,
width=\figurewidth,
height=\figureheight,
tick align=outside,
tick pos=left,
x grid style={white!69.01960784313725!black},
y grid style={white!69.01960784313725!black},
legend entries={{10 peers},{20 peers},{30 peers}},
legend style={at={(0.97,0.03)}, anchor=south east, draw=white!80.0!black},
legend cell align={left}
]
\addplot [red, mark=*, mark size=1, mark options={solid}]
table {%
1 0
2 0
4 0
6 0
8 0
10 0
12 0
14 0
16 0
18 0
20 18
30 101
40 101
50 101
55 101
60 101
65 101
70 101
80 101
90 101
100 101
};
\addplot [green!50.0!black, mark=square*, mark size=1, mark options={solid}]
table {%
1 0
2 0
4 0
6 0
8 0
10 88
12 101
14 100
16 101
18 101
20 101
30 101
40 101
50 101
55 101
60 101
65 101
70 101
80 101
90 101
100 101
};
\addplot [blue, mark=diamond*, mark size=1, mark options={solid}]
table {%
1 0
2 0
4 0
6 75
8 101
10 101
12 101
14 101
16 101
18 101
20 101
30 101
40 101
50 101
55 101
60 101
65 101
70 101
80 101
90 103
100 101
};
\end{axis}

\end{tikzpicture}
\vspace{-6mm}
  \caption{Computation Request Protocol: Saturation of the uwsgi request queue depending on the amount of requests performed by the client. During each request, the gateway contacts peers and waits for their response.  The longer time needed by each request already causes the queue to fill up when requests come in with a delay of 100 to 200\,ms.}
  \label{fig:ir_gateway_queue}
\end{figure}
\fi
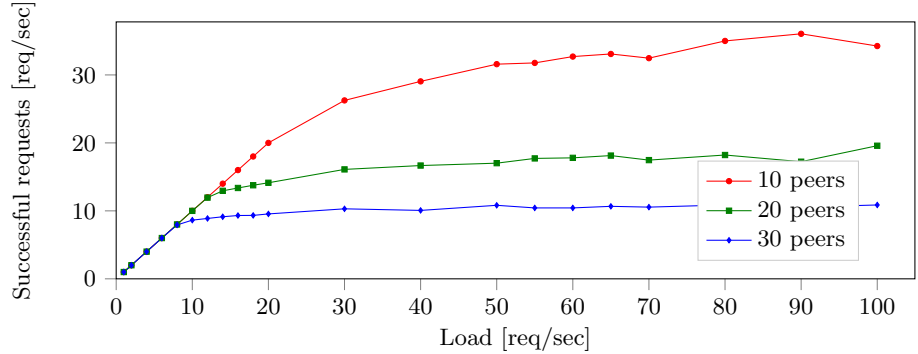
\begin{figure}[t]
\begin{tikzpicture}

\begin{axis}[
xlabel={Load [req/sec]},
ylabel={Successful requests [req/sec]},
xmin=0, xmax=104.95,
ymin=0, ymax=37.8025,
width=\figurewidth,
height=\figureheight,
tick align=outside,
tick pos=left,
x grid style={white!69.01960784313725!black},
y grid style={white!69.01960784313725!black},
legend entries={{10 peers},{20 peers},{30 peers}},
legend style={at={(0.93,0.07)}, anchor=south east, draw=white!80.0!black},
legend cell align={left}
]
\addplot [red, mark=*, mark size=1, mark options={solid}]
table {%
1 1
2 2
4 4
6 6
8 8
10 10
12 12
14 14
16 16
18 18
20 20
30 26.25
40 29.05
50 31.5833333333333
55 31.7666666666667
60 32.7
65 33.0833333333333
70 32.4666666666667
80 35
90 36.05
100 34.25
};
\addplot [green!50.0!black, mark=square*, mark size=1, mark options={solid}]
table {%
1 1
2 2
4 4
6 6
8 8
10 10
12 11.95
14 12.95
16 13.3666666666667
18 13.7666666666667
20 14.1333333333333
30 16.1
40 16.6666666666667
50 17.0166666666667
55 17.7166666666667
60 17.8
65 18.1333333333333
70 17.4666666666667
80 18.2166666666667
90 17.2333333333333
100 19.5833333333333
};
\addplot [blue, mark=diamond*, mark size=1, mark options={solid}]
table {%
1 1
2 2
4 4
6 6
8 7.93333333333333
10 8.63333333333333
12 8.88333333333333
14 9.13333333333333
16 9.31666666666667
18 9.33333333333333
20 9.55
30 10.3
40 10.0666666666667
50 10.8166666666667
55 10.4333333333333
60 10.4333333333333
65 10.6666666666667
70 10.55
80 10.8666666666667
90 10.55
100 10.8666666666667
};
\end{axis}

\end{tikzpicture}
\vspace{-6mm}
  \caption{Computation Request Protocol: The amount of successfully answered requests depending on the amount of requests performed by the client. Latency of single requests restricts the amount of successful requests, since the request queue is already filled between 5 -- 10 requests/s.}
  \label{fig:ir_gateway_throughput_p1}
\end{figure}

Concomitant with the increase in latency, the request queue is exhausted early between 5 -- 20 requests/s\iflongversion{} (Figure~\ref{fig:ir_gateway_queue})\fi{} and a high throughput is inhibited (Figure~\ref{fig:ir_gateway_throughput_p1}).
Here, the bottleneck is the CPU of the gateway.
Providing 4 uwsgi processes shows that parallelization doubles the throughput\iflongversion{} (Figure~\ref{fig:ir_gateway_throughput_p4})\fi.
\iflongversion
\begin{figure}[t]
\begin{tikzpicture}

\begin{axis}[
xlabel={Load [req/sec]},
ylabel={Successful requests [req/sec]},
xmin=0, xmax=104.95,
ymin=0, ymax=83.46,
width=\figurewidth,
height=\figureheight,
tick align=outside,
tick pos=left,
x grid style={white!69.01960784313725!black},
y grid style={white!69.01960784313725!black},
legend entries={{10 peers},{20 peers},{30 peers}},
legend style={at={(0.03,0.97)}, anchor=north west, draw=white!80.0!black},
legend cell align={left}
]
\addplot [red, mark=*, mark size=1, mark options={solid}]
table {%
1 1
2 2
4 4
6 6
8 8
10 10
12 12
14 14
16 16
18 18
20 20
30 30
40 40
50 50
55 55
60 60
65 63.2666666666667
70 65.85
80 71.0833333333333
90 73.9666666666667
100 79.5333333333333
};
\addplot [green!50.0!black, mark=square*, mark size=1, mark options={solid}]
table {%
1 1
2 2
4 4
6 6
8 8
10 10
12 12
14 14
16 16
18 18
20 20
30 29.3666666666667
40 33.3
50 36.25
55 37.5833333333333
60 38.3833333333333
65 39.6166666666667
70 40.6
80 40.65
90 43.0166666666667
100 44.3333333333333
};
\addplot [blue, mark=diamond*, mark size=1, mark options={solid}]
table {%
1 1
2 2
4 4
6 6
8 8
10 10
12 12
14 14
16 16
18 17.7166666666667
20 18.7833333333333
30 21.1333333333333
40 24.9166666666667
50 25.65
55 25.6333333333333
60 27.1833333333333
65 26.5666666666667
70 26.5333333333333
80 27.8333333333333
90 28.0166666666667
100 28.6166666666667
};
\end{axis}

\end{tikzpicture}
\vspace{-6mm}
\caption{Computation Request Protocol: The amount of successfully answered requests depending on the amount of requests performed by the client. A higher amount of gateway processes improves throughput correspondingly.  }
  \label{fig:ir_gateway_throughput_p4}
\end{figure}
\fi

\section{Summary}
\label{sec:summary}
In the Internet of Things and smart environments, services need data about the environment and its inhabitants for action.
\iflongversion
Since this data also provides insights into user presence and behavior,
\else
However,
\fi
this data is personal data and, hence,  privacy critical.

The state of the art handles this data by deploying a middleware for storage and processing.
The results are then forwarded to services and applications.
However, this centralization enables several privacy threats.
Hence, a more privacy-preserving method of data processing should be considered.

A promising approach is Secure Multiparty Computation (SMC).
Data can remain on individual distributed sensor platforms.
Afterwards, desired results can be derived on-the-fly by secure computation without intermediate storage.

We provide a method to execute queries on privacy-sensitive sensor data by extending SMC:
Our approach allows third party clients to request data processing from SMC without taking part in the computations themselves.
Furthermore, we add further elements which are needed for obtaining a fully privacy-preserving service:
Protection of results is achieved by access control on the level of individual queries.
Requests and processing becomes fully transparent for the cooperating sensor platforms.
This supports intervenability, \ie enabling  them to stay in control of their data and
and allowing them to reject processing if their privacy requirements for incoming data requests are not fulfilled.
\paragraph*{Future Work}
\label{sec:future_work}
In combination with \cite{vonMaltitz2018b}, we realized the full architecture of a privacy-preserving service for smart environments.
However, several further challenges are open.
\inlineEnum{remaining}Abstractly, we enabled checking of requests by peers given local policies $\Phi_p$.
The design of these semantic checks could be further investigated.
\inlineEnum{remaining}Clients are only allowed to select predefined predicates due to privacy reasons.
If arbitrary predicates could be checked against privacy rules,
clients could be allowed to define predicates themselves.
\inlineEnum{remaining}The gateway could create its access control structure $\Phi$ by collecting and merging the distributed set of $\Phi_p$, directly guaranteeing compatibility between them.
\iflongversion
\inlineEnum{remaining}Lastly, distributed consistent accountability could be supported by adding a distributed ledger technology acting as an append-only medium.
This approach is realistic, since the premise to have distributed peers of different owners is already given in this context.
\fi

\bibliography{library}

\bibliographystyle{splncs04}
\end{document}